\documentclass[preprint,aps,pra,showpacs,floatfix]{revtex4}
\usepackage{amsmath}
\usepackage{times}
\usepackage {euscript}
\usepackage{graphicx}
\usepackage{amsthm}
\usepackage{epsfig}
\usepackage{bm}
\usepackage{multirow}
\begin{document}
\thispagestyle{empty}
%%%%%%%%%%­%%%%%%%%%%­%%%%%%%%%%­%%%%%%%%%%­%%%%%%%%%%­%%%%%%%%%%­%%%%%%%%%%%%
\title{
Parity nonconservation effect in the dielectronic recombination of 
polarized electrons with heavy He-like ions
}
\author{V.~A.~Zaytsev$^{1}$,
        A.~V.~Maiorova$^{1}$,
        V.~M.~Shabaev$^{1}$,
        A.~V.~Volotka$^{1,2}$,
        S.~Tashenov$^{3}$,
        G.~Plunien$^{2}$,
        and Th.~St\"ohlker$^{4,5,6}$}
\affiliation{
$^1$ Department of Physics, St. Petersburg State University,
Ulianovskaya 1, Petrodvorets, 198504 St. Petersburg, Russia \\
$^2$ Institut f\"ur Theoretische Physik, Technische Universit\"at 
Dresden, Mommsenstra{\ss}e 13, D-01062 Dresden, Germany\\
$^3$ Physikalisches Institut, Universit\"at Heidelberg, D-69120 
Heidelberg, Germany\\
$^4$ GSI Helmholtzzentrum f\"ur Schwerionenforschung GmbH, D-64291 
Darmstadt , Germany\\
$^5$ Helmholtz-Institut Jena, D-07743 Jena, Germany\\
$^6$ Institut f\"ur Optik und Quantenelektronik, 
Friedrich-Schiller-Universit\"at Jena, D-07743 Jena, Germany
\vspace{10mm}
}
%%%%%%%%%%­%%%%%%%%%%­%%%%%%%%%%­%%%%%%%%%%­%%%%%%%%%%­%%%%%%%%%%­%%%%%%%%%%%%
%
%=========­==========­===       ABSTRACT         ==========­==========­=====
\begin{abstract}
% abstract - I.1
We investigate the parity nonconservation (PNC) effect in the 
dielectronic recombination (DR) of a polarized
electron with a heavy He-like ion into doubly-excited $\left(
\left( 1s 2p_{1/2} \right)_{0} n\kappa \right)_{1/2}$ and $\left(
\left( 1s 2s \right)_{0} n\kappa \right)_{1/2}$ states of Li-like ion.
% abstract - I.2
We determine the nuclear charge number $Z$ for
which these opposite-parity levels are near to cross and, therefore,
the PNC effect will be significantly enhanced.
% abstract - I.3
Calculations are performed for quantum numbers $n \geq 4$ and
$\kappa = \pm 1$.
\end{abstract}
%=========­==========­===       ABSTRACT         ==========­==========­=====
%
%%%%%%%%%%­%%%%%%%%%%­%%%%%%%%%%­%%%%%%%%%%­%%%%%%%%%%­%%%%%%%%%%­%%%%%%%%%%­%%
\maketitle
%%%%%%%%%%­%%%%%%%%%%­%%%%%%%%%%­%%%%%%%%%%­%%%%%%%%%%­%%%%%%%%%%­%%%%%%%%%%­%%
\section{INTRODUCTION}
%=========­==========­===      INTRODUCTION         ==========­==========­====
% ---------------------       I paragraph         ------------------------
% Introduction - I.1
Investigations of the parity nonconservation (PNC) effects in
atoms play a very important role for tests of the electroweak sector of
the Standard Model (SM) in the low-energy regime~\cite{Khriplovich,Khriplovich_PST112_52:2004,Ginges_PR397_63:2004}.
% Introduction - I.2
The unprecedented experimental precision for the PNC amplitude was
obtained in $^{133}$Cs measurements~\cite{Wood_S275_1759:1997,Bennett_PRL82_2484:1999}
and, together with recent progress in QED and atomic structure
calculations (see, e.g., Refs.~\cite{Shabaev_PRL94_213002:2005,
Porsev_PRL102_181601:2009, Dzuba_PRL109_203003:2012} and references
therein), provided the most accurate to date test of the SM with atomic
systems.
% Introduction - I.3
From the theoretical side, further progress in studying the PNC
effect with neutral atoms is strongly limited by the uncertainties of
the electron-correlation contributions.
% Introduction - I.4
In contrast to that, in heavy highly-charged ions the correlation effects,
being suppressed by a factor $1/Z$, can be calculated by perturbation
theory up to the required precision.
% Introduction - I.5
This gives good prospects for studying the PNC effects with highly-charged ions.
%-----------------------------------------------------------------------

% ---------------------       II paragraph        ------------------------
% Introduction - II.1
PNC experiments with few-electron ions were first proposed by Gorshkov
and Labzowsky in Ref.~\cite{Gorshkov_JL19_394:1974}, where the fact that
opposite-parity 2$^{1}S_{0}$ and 2$^{3}P_{1}$ states are near to cross
for He-like ions with $Z \sim 6$ and $Z \sim 29$ was utilized.
% Introduction - II.2
Since that work, a number of authors considered He-like ions as very
promising systems for investigating the PNC
effects~\cite{Schafer_PRA40_7362:1989,
Pindzola_PRA47_4856:1993,dun96,Labzowsky_PRA63_054105:2001,
Gribakin_PRA72_032109:2005, Maiorova_JPB42_205002:2009,
Shabaev_PRA81_052102:2010, Ferro_PRA81_062503:2010,
Ferro_PRA83_052518:2011, Maiorova_JPB44_225003:2011,
Gunst_PRA87_032714:2013}.
% Introduction - II.3
This is due to the fact that
%He-like ions have a relatively simple electronic structure.
% Introduction - II.4
%Moreover,
the PNC effects in He-like ions can be significantly
enhanced due to the near-degeneracy of some opposite-parity states.
% Introduction - II.5
In a large number of proposals~\cite{Schafer_PRA40_7362:1989,
Labzowsky_PRA63_054105:2001,
Maiorova_JPB42_205002:2009, Shabaev_PRA81_052102:2010,
Ferro_PRA81_062503:2010, Ferro_PRA83_052518:2011,
Maiorova_JPB44_225003:2011, Gunst_PRA87_032714:2013} the level crossing
between the $2^3P_0$ and $2^1S_0$ states of He-like ions was exploited.
% Introduction - II.6
%With this in mind,
One may expect that the addition of a highly-excited
electron would not strongly change the energy difference between the
corresponding levels in Li-like ions.
% Introduction - II.7
Indeed, the opposite-parity $\left( \left(1s 2s\right)_{0} n\kappa
\right)_{1/2}$ and $\left( \left(1s 2p_{1/2}\right)_{0} n\kappa
\right)_{1/2}$ states can be still made almost to cross by choosing the
principal quantum number $n$ and the Dirac angular quantum number
$\kappa = (-1)^{j+l+1/2}(j +1/2)$.
% Introduction - II.8
In this work we present such quasidegenerate levels of heavy
Li-like ions and propose a scheme for observing the PNC effect in
dielectronic recombination (DR) of free electrons with He-like
ions into these double-excited states of Li-like ions.
%-----------------------------------------------------------------------

%----------------------        III paragraph             ------------------
% Introduction - III.1
In some previous proposals the dielectronic recombination was
considered as a convenient probe process, which can be
used to measure the parity violation effects.
% Introduction - III.2
In Ref.~\cite{Pindzola_PRA47_4856:1993}, Pindzola studied the PNC
effect on the Auger-electron emission from He-like uranium.
% Introduction - III.3
The parity violation in dielectronic recombination of polarized
electrons with H-like ions at $Z < 60$ was discussed by Gribakin \textit{et al.} in Ref.~\cite{Gribakin_PRA72_032109:2005}.
% Introduction - III.4
In our previous work~\cite{Maiorova_JPB44_225003:2011} we
investigated the PNC effect on recombination of a polarized
electron with unpolarized H-like thorium ($Z = 90$) and gadolinium
($Z = 64$) ions in the case of resonance with a doubly-excited
state of the corresponding He-like ions.
% Introduction - III.5
In the present work we investigate the PNC effect in the process of the
dielectronic recombination of polarized electrons with heavy
He-like ions into the doubly-excited $\left( \left(1s
2s\right)_{0} n\kappa \right)_{1/2}$ and $\left( \left(1s
2p_{1/2}\right)_{0} n\kappa \right)_{1/2}$ states of Li-like ions.
% Introduction - III.6
The energy of the incident electron is considered
to be tuned in resonance with one of these levels.
% Introduction - III.7
The case of non-monoenergetic incident electron beam is also studied.
%-----------------------------------------------------------------------

%---------------------          IV paragraph          --------------------
% Introduction - IV.1
Throughout the paper relativistic units ($\hbar = c = 1$) and Heaviside charge unit ($\alpha = e^{2}/(4\pi), e < 0$) are used.
%=========­==========­==         END INTRODUCTION       ==========­==========­

%=========­==========­==         BASIC FORMALISM          ==========­========
\section{BASIC FORMALISM}
\label{sec:BF}
%-----------------------              I paragraph        ------------------
% Basic formalism - I.1
We consider the process of the dielectronic recombination of an electron
having asymptotic four-momentum $p_{i} = \left(\varepsilon_{i}, \mathbf{p}_{i} \right)$
and polarization $\mu_{i}$ with a heavy He-like ion, being originally
in the ground $(1s)^2$ state.
% Basic formalism - I.2
As a result of this non-radiative capture, one of the near-degenerate 
opposite-parity $d_{1}$ or $d_{2}$ states of the Li-like ion is formed.
% Basic formalism - I.3
To simplify the derivation of formulas, we assume that these
levels decay via the emission of a photon to some final state $f$.
% Basic formalism - I.4
We suppose that the incoming electron energy $\varepsilon_{i}$ is
chosen to get the resonance with one of the doubly-excited
$d_1$ or $d_2$ states.
% Basic formalism - I.5
The differential cross section of the process under consideration is
defined as~\cite{Shabaev_PRA50_4521:1994,Shabaev_PR356_119:2002}
\begin{eqnarray}
\nonumber
\frac{d\sigma_{\mu_{i}}}{d\Omega} & = & \frac{\left(2\pi\right)^4}{v_{i}}\omega^{2} \sum_{\boldsymbol{\epsilon}_{f}}
\left| \sum_{M_{d_{1}}} \tau_{\gamma_{f},f;d_{1}} \frac{1}{E_{i} - E_{d_{1}} + \text{i} \Gamma_{d_{1}}/2}   \left\langle \Psi_{d_{1}}|I| \Psi_{i} \right\rangle \right.
\\
&& \left. + \sum_{M_{d_{2}}} \tau_{\gamma_{f},f;d_{2}}
\frac{1}{E_{i} - E_{d_{2}} + \text{i}\Gamma_{d_{2}}/2}
\left\langle \Psi_{d_{2}}|I| \Psi_{i} \right\rangle \right|^{2},
\label{eq:partial_sigma}
\end{eqnarray}
where $E_{d_{k}}$, $\Gamma_{d_{k}}$, and $M_{d_{k}}$ are the energy,
the total width, and the momentum projection of the $d_{k}$ state
($k = 1,2$), respectively.
% Basic formalism - I.6
$E_{i} = E_{\left(1s\right)^{2}} + \varepsilon_{i}$ is the total energy
of the initial state of the system and $v_{i}$ is the velocity
of the incident electron.
% Basic formalism - I.7
The outgoing photon $\gamma_{f}$ is characterized by the energy
$\omega$ and the polarization $\boldsymbol{\epsilon}_{f}$.
% Basic formalism - I.8
$\tau_{\gamma_{f},f;d_k}$ is the amplitude of the radiative
transition from the $d_k$ state to the $f$ state via the emission of a
photon and $I$ is the operator of the interelectronic interaction as
defined in Ref.~\cite{Shabaev_PR356_119:2002}.
%-----------------------------------------------------------------------

%-------------------            II paragraph        -----------------------
% Basic formalism - II.1
As mentioned above, for heavy few-electron ions the
interelectronic-interaction effects are suppressed by a factor
$1/Z$, compared to the interaction of the electrons with the Coulomb
field of the nucleus.
% Basic formalism - II.2
Therefore, we can generally consider the wave functions of our system
in the independent-electron approximation.
% Basic formalism - II.3
With this approximation, the initial state wave function is given by
\begin{eqnarray}
\nonumber
\Psi_{p_{i}\mu_{i},JM} \left(
\mathbf{x}_{1},\mathbf{x}_{2},\mathbf{x}_{3} \right)
&=& A_{N} \sum_{\mathcal{P}} (-1)^{\mathcal{P}} \mathcal{P} \sum_{m_{1}m_{2}}
C_{j_{1}m_{1}, j_{2}m_{2}}^{JM}
\\
&& \times
\psi_{n_{1}\kappa_{1}m_{1}}\left(\mathbf{x}_{1}\right)
\psi_{n_{2}\kappa_{2}m_{2}}\left(\mathbf{x}_{2}\right)
\psi_{p_{i}\mu_{i}}\left(\mathbf{x}_{3}\right),
\label{eq:li_wf_cd}
\end{eqnarray}
where $\psi_{n\kappa m}\left(\mathbf{x}\right)$ is the one-electron
bound-state Dirac wave function, $\psi_{p_{i}\mu_{i}}\left(\mathbf{x}\right)$ is the
incident electron wave function, $C_{j_{1}m_{1}, j_{2}m_{2}}^{JM}$
is the Clebsch-Gordan coefficient, $\left(-1\right)^{\mathcal{P}}$
is the parity of the permutation, $\mathcal{P}$ is the permutation
operator, and $A_{N}$ is the normalization factor.
% Basic formalism - II.4
From the theoretical viewpoint, it is convenient to formulate the
electron capture in the ion rest frame.
% Basic formalism - II.5
In this frame we can adopt that the quantization axis
($z$ - axis) is directed along the incoming electron momentum $\mathbf{p}_{i}$.
% Basic formalism - II.6
In this case the full expansion of the incoming electron wave
function is given by (see, e.g., Refs.~\cite{Eichler_PR439_1:2007,Rose})
\begin{equation}
\psi_{p_{i}\mu_{i}}\left(\mathbf{x}\right) =
\frac{1}{\sqrt{4\pi}}\frac{1}{\sqrt{p_{i}\varepsilon_{i}}}\sum_{\kappa}
\text{i}^{l} \exp(\text{i}\Delta_{\kappa}) \sqrt{2l + 1} C_{l0,
1/2\mu_{i}}^{j\mu_{i}}
\psi_{\epsilon_{i}\kappa\mu_{i}}\left(\mathbf{x}\right),
\label{eq:wf_free_electron}
\end{equation}
%
%
% Basic formalism - II.7
where $\Delta_{\kappa}$ is the Coulomb phase shift,
$\psi_{\varepsilon_{i}\kappa\mu_{i}}\left(\mathbf{x}\right)$ is the
partial electron wave with the Dirac quantum number
$\kappa = (-1)^{j+l+1/2}(j +1/2)$, determined
by the angular momentum $j$ and the parity $l$.
%-----------------------------------------------------------------------

%-------------------             III paragraph     ------------------------
% Basic formalism - III.1
Neglecting the weak interaction, we can write the wave functions
of the intermediate $d$ and final $f$ states as follows
\begin{eqnarray}
\nonumber \Psi_{J(J')M} \left(
\mathbf{x}_{1},\mathbf{x}_{2},\mathbf{x}_{3} \right)
& = & A_{N} \sum_{\mathcal{P}} (-1)^{\mathcal{P}} \mathcal{P} \sum_{M'm_{3}}
\sum_{m_{1}m_{2}} C_{J'M', j_{3}m_{3}}^{JM} C_{j_{1}m_{1},
j_{2}m_{2}}^{J'M'}
\\
&& \times
\psi_{n_{1}\kappa_{1}m_{1}}\left(\mathbf{x}_{1}\right)
\psi_{n_{2}\kappa_{2}m_{2}}\left(\mathbf{x}_{2}\right)
\psi_{n_{3}\kappa_{3}m_{3}}\left(\mathbf{x}_{3}\right).
\label{eq:li_wf_dd}
\end{eqnarray}
%
%
% Basic formalism - III.2
To account for the weak interaction, the intermediate $d_1$ and $d_2$
states should be considered with a small admixture of the
closest-lying, opposite-parity $d_2$ and $d_1$ states, respectively.
% Basic formalism - III.3
Then, the wave functions of the corresponding doubly-excited states modify as
\begin{eqnarray}
\left| \Psi_{d_{1}} \right\rangle \rightarrow \left| \Psi_{d_{1}}
\right\rangle + \text{i} \xi \left| \Psi_{d_{2}} \right\rangle,
\\
\left| \Psi_{d_{2}} \right\rangle \rightarrow \left| \Psi_{d_{2}}
\right\rangle + \text{i} \xi \left| \Psi_{d_{1}} \right\rangle,
\end{eqnarray}
%
%
% Basic formalism - III.4
where the admixing parameter $\text{i}\xi = \left\langle
\Psi_{d_{2}} \left| \sum_{i=1}^{3} H_{\text{W}}\left(i\right) \right| \Psi_{d_{1}}
\right\rangle / \left(E_{d_{1}} - E_{d_{2}}\right)$ is
determined by the nuclear spin-independent effective Hamiltonian of weak
interaction
\begin{equation}
H_{\text{W}} = -\left(G_{\text{F}} / \sqrt{8} \right) Q_{\text{W}} \rho_{\text{N}}\left( r \right) \gamma_{5}.
\label{eq:weak_hamiltonian}
\end{equation}
Here $Q_{\text{W}} \approx -N + Z\left(1 - 4
\sin^{2}\theta_{\text{W}}\right)$ denotes the weak charge of the
nucleus, $G_{\text{F}}$ is the Fermi constant, $\gamma_{5}$ is the
Dirac matrix, and $\rho_{\text{N}}$ is the nuclear weak-charge
density (normalized to unity).
% Basic formalism - III.5
After substitution of the modified wave functions into
Eq.~(\ref{eq:partial_sigma}) and summing over all decay channels,
one finds
\begin{eqnarray}
\nonumber \sigma_{\mu_{i}} & = &
\frac{\left(2\pi\right)^{3}}{v_{i}}
\sum_{M_{d_{1}}M_{d_{2}}}
\delta_{M_{d_{1}}M_{d_{2}}}
\left[
\frac{\Gamma_{d_{1}}}{\left|
E_{i} - E_{d_{1}} + \text{i} \Gamma_{d_{1}}/2 \right|^{2}} \left|
\left\langle \Psi_{d_{1}} |I| \Psi_{i}\right\rangle \right|^{2}
+
\frac{\Gamma_{d_{2}}}{\left| E_{i} - E_{d_{2}} + \text{i}
\Gamma_{d_{2}}/2 \right|^{2}} \left|
\left\langle \Psi_{d_{2}} |I| \Psi_{i}\right\rangle \right|^{2}
\right.
\\ \nonumber
&& + 2 \left(
\frac{\Gamma_{d_{1}}}{\left| E_{i} - E_{d_{1}} +
\text{i} \Gamma_{d_{1}}/2 \right|^{2}}
-
\frac{\Gamma_{d_{2}}}{\left| E_{i} - E_{d_{2}} + \text{i}
\Gamma_{d_{2}}/2 \right|^{2}}
\right)
\Re \left(\text{i} \xi
\left\langle \Psi_{d_{1}} |I| \Psi_{i} \right\rangle
\left(\left\langle \Psi_{d_{2}} |I| \Psi_{i}\right\rangle\right)^{*}\right)
\\
&& \left.
+ 2 \left(\Gamma_{d_{2}} - \Gamma_{d_{1}} \right)
\Re
\left( \text{i} \xi
\frac{\left\langle \Psi_{d_{1}} |I| \Psi_{i} \right\rangle
\left(\left\langle \Psi_{d_{2}}|I| \Psi_{i} \right\rangle\right)^{*}}{
\left(E_{i} - E_{d_{1}} + \text{i}\Gamma_{d_{1}}/2\right)
\left(E_{i} - E_{d_{2}} - \text{i}\Gamma_{d_{2}}/2\right)}
\right)
\right].
\label{eq:sigma_pnc}
\end{eqnarray}
%
%
% Basic formalism - III.6
In this expression the terms of order $\xi^{2}$ are neglected.
% Basic formalism - III.7
The first and the second terms are parity conserving, while the
third and the fourth terms correspond to the P-violating
contributions to the cross section.
% Basic formalism - III.8
The third term originates from the weak interaction in the dielectronic recombination process.
% Basic formalism - III.9
The P-violation in the decay process is described by the fourth term.
% Basic formalism - III.10
In the case of clearly resolved levels ($\Gamma_{d_{1}}, \Gamma_{d_{2}} \ll \left|
E_{d_{1}} - E_{d_{2}}\right|$), one can consider only the resonant term
in Eq.~(\ref{eq:partial_sigma}).
% Basic formalism - III.11
For example, if the energy of the incident electron is tuned to the
$d_{1}$ state, the total cross section takes a form
\begin{eqnarray}
\sigma_{\mu_{i}} = \frac{\left(2\pi\right)^{3}}{v_{i}}
\frac{\Gamma_{d_{1}}}{\left| E_{i} - E_{d_{1}} + \text{i}
\Gamma_{d_{1}}/2 \right|^{2}} \sum_{M_{d_{1}}M_{d_{2}}}
\delta_{M_{d_{1}}M_{d_{2}}}
\left[
\left|\left\langle \Psi_{d_{1}} |I| \Psi_{i} \right\rangle \right|^{2}
+ 2 \Re
\left(
\text{i} \xi
\left\langle \Psi_{d_{1}} |I| \Psi_{i} \right\rangle
\left(\left\langle \Psi_{d_{2}} |I| \Psi_{i} \right\rangle\right)^{*}
\right)
\right].
\label{eq:sigma_pnc_res}
\end{eqnarray}
%
%
% ----------------------------------------------------------------------

%--------------------             IV paragraph     ------------------------
% Basic formalism - IV.1
In the case when the energy spread of the electron beam exceeds
the energy spacing between the quasidegenerate states, one should
integrate Eq.~(\ref{eq:sigma_pnc}) over the incident electron energies.
% Basic formalism - IV.2
It can be performed analytically since the velocity $v_{i}$ and the DR
amplitudes weakly change within the interval of the beam energy distribution.
% Basic formalism - IV.3
Thus, for the close lying states, one obtains
\begin{eqnarray}
\nonumber \overline{\sigma}_{\mu_{i}} &=& \frac{\left(2\pi\right)^{4}}{v_{i}}
\sum_{M_{d_{1}}M_{d_{2}}}
\delta_{M_{d_{1}}M_{d_{2}}} \biggl\{
\left| \left\langle \Psi_{d_{1}} |I| \Psi_{i} \right\rangle \right|^{2} +
\left| \left\langle \Psi_{d_{2}} |I| \Psi_{i} \right\rangle \right|^{2}
\\
&& \left.
+ 2 \xi \left(\Gamma_{d_{2}} - \Gamma_{d_{1}}\right) \Re
\left[
\frac{\left\langle \Psi_{d_{1}} |I| \Psi_{i} \right\rangle
\left(\left\langle \Psi_{d_{2}} |I| \Psi_{i} \right\rangle\right)^{*}}{\left(E_{d_{1}}
- E_{d_{2}}\right) - \text{i} \left(\Gamma_{d_{1}} +
\Gamma_{d_{2}}\right)/2} \right] \right\rbrace,
\label{eq:sigma_pnc_int}
\end{eqnarray}
where $\overline{\sigma}_{\mu_{i}}$ is the integrated cross section.
% Basic formalism - IV.4
When the energy distribution in the beam exceeds the energy widths but is much less than the energy distance between the quasidegenerate levels, one should integrate Eq.~(\ref{eq:sigma_pnc_res}).   
% Basic formalism - IV.5
For instance, for a non-monoenergetic beam tuned to the $d_{1}$ state we obtain
\begin{eqnarray}
\overline{\sigma}_{\mu_{i}} = \frac{\left(2\pi\right)^{4}}{v_{i}}
\sum_{M_{d_{1}}M_{d_{2}}}
\delta_{M_{d_{1}}M_{d_{2}}}
\left[
\left| \left\langle \Psi_{d_{1}} |I| \Psi_{i} \right\rangle\right|^{2}
+ 2 \Re
\left(\text{i}\xi\left\langle \Psi_{d_{1}} |I| \Psi_{i} \right\rangle
\left(\left\langle \Psi_{d_{2}} |I| \Psi_{i} \right\rangle\right)^{*}\right)
\right].
\label{eq:sigma_pnc_res_int}
\end{eqnarray}
%-----------------------------------------------------------------------
%=========­==========­===        END BASIC FORMALISM             ==========­==

%=========­==========­==         RESULTS AND DISCUSSION         ==========­==
\section{RESULTS AND DISCUSSION}
\label{sec:RD}
%------------------             I paragraph        ------------------------
% Results and discussion - I.1
As mentioned above, the enhancement of the PNC effect takes place for close-lying opposite-parity levels.
% Results and discussion - I.2
In our previous work~\cite{Zaytsev_PS} we found that for Li-like
ions the near degeneracy takes place for several doubly-excited
opposite-parity $\left( \left(1s 2s\right)_{0} n\kappa
\right)_{1/2}$ and $\left( \left(1s 2p_{1/2}\right)_{0} n\kappa
\right)_{1/2}$ states with $4 \leq n \leq 7$, $\kappa = \pm 1$, $Z
\sim 60$, and $Z \sim 92$.
% Results and discussion - I.3
The energy difference has been evaluated as follows
\begin{equation}
\Delta E = E_{\left( \left(1s 2p_{1/2}\right)_{0} n\kappa \right)_{1/2}} - E_{\left( \left(1s 2s\right)_{0} n\kappa \right)_{1/2}} = \Delta E^{(\text{He})} +
\Delta E^{(\text{Ext})},
\end{equation}
where $\Delta E^{(\text{He})} = E_{\left(1s 2p_{1/2}\right)_{0}} -
E_{\left(1s 2s\right)_{0}}$ is the energy difference of the corresponding
levels in He-like ion and $\Delta E^{(\text{Ext})} =
E_{\left(1s 2p_{1/2}\right)_{0}}^{\left(n\kappa\right)} -
E_{\left(1s 2s\right)_{0}}^{\left(n\kappa\right)}$ is the difference of the one-photon exchange contributions, describing the interaction between the external $n \kappa$ electron and
the inner-shell electrons.
% Results and discussion - I.4
The highly accurate values of $\Delta E^{(\text{He})}$, including all second-order two-electron QED contributions, were taken from Ref.~\cite{Artemyev_PRA71_062104:2005}.
% Results and discussion - I.5
We also have taken into account the mixing of the close-lying
$\left(1s 2s ns\right)_{1/2}$ and $\left(1s 2p_{1/2} np_{1/2}\right)_{1/2}$
levels, as well as the $\left(1s 2p_{1/2} ns\right)_{1/2}$ and
$\left(1s 2s np_{1/2}\right)_{1/2}$ levels (see Ref.~\cite{Zaytsev_PS} for details).
% ----------------------------------------------------------------------

%------------------------      II paragraph         ----------------------
% Results and discussion - II.1
In the present work we consider the PNC effect in the process of the
dielectronic recombination into $d_{1} = \left(\left(1s 2p_{1/2}\right)_{0}n\kappa\right)_{1/2}$ and $d_{2} = \left(\left(1s 2s\right)_{0}n\kappa\right)_{1/2}$ states of Li-like ions.
% Results and discussion - II.2
First, let us denote the cross sections for positive and negative
helicities (spin projection onto the electron momentum direction)
of the incident electron as $\sigma_{+}$ and $\sigma_{-}$,
respectively.
% Results and discussion - II.3
We also introduce designations for the cross section without the
PNC effect, $\sigma_{0} = \left(\sigma_{+} + \sigma_{-}\right)/2$,
and the PNC contribution, $\sigma_{\text{PNC}} = \left(\sigma_{+}
- \sigma_{-}\right)/2$.
% Results and discussion - II.4
Deviation of $\sigma_{\text{PNC}}$ from zero indicates the parity violation effect.
% Results and discussion - II.5
Finally, one should determine the requirements on the luminosity $L$,
provided the PNC effect is measured to a relative accuracy
$\eta$~\cite{Gribakin_PRA72_032109:2005,
Maiorova_JPB42_205002:2009}
\begin{equation}
L > L_{0} = \frac{\sigma_{+} + \sigma_{-} +
2\sigma_{\text{b}}}{\left(\sigma_{+} -
\sigma_{-}\right)^{2}\eta^{2}T}.
\label{eq:luminosity}
\end{equation}
%
%
% Results and discussion - II.6
Here $\sigma_{\text{b}}$ is the background magnitude and $T$ is
the acquisition time.
% Results and discussion - II.7
In our calculations we neglect the background signal, set $T =
$ 2 weeks, and $\eta = 0.01$.
% Results and discussion - II.8
In the case of non-monoenergetic incident electron beam the integrated cross
sections $\overline{\sigma}_{0} = \left(\overline{\sigma}_{+} + \overline{\sigma}_{-}\right)/2$ and
$\overline{\sigma}_{\text{PNC}} = \left(\overline{\sigma}_{+} - \overline{\sigma}_{-}\right)/2$
should be used instead of $\sigma_{0}$ and $\sigma_{\text{PNC}}$.
% Results and discussion - II.9
Here we denote the integrated cross sections for
positive and negative   helicities of the incident electron as
$\overline{\sigma}_{+}$ and $\overline{\sigma}_{-}$, respectively.
% ----------------------------------------------------------------------

%---------------------          III paragraph            ---------------
% Results and discussion - III.1
In order to investigate either the levels mixed by the weak
interaction are distinguished or not, we introduce the coefficient
$R = \left|E_{d_{1}} - E_{d_{2}}\right| / \left(\Gamma_{d_{1}} +
\Gamma_{d_{2}}\right)$.
% Results and discussion - III.2
Evaluating the cross section according to
Eqs.~(\ref{eq:sigma_pnc}) and~(\ref{eq:sigma_pnc_res}), it was
found that the results became similar at $R \geq 5$.
% Results and discussion - III.3
Thus, levels with $R \geq 5$ are regarded as distinguishable.
% ----------------------------------------------------------------------

%---------------------          IV paragraph      ----------------------
% Results and discussion - IV.1
In Tables~\ref{tb:minimum_l0_res} and \ref{tb:minimum_l0} we
present numerical results for the most promising case of the
resonance DR into the
$\left(\left(1s2p_{1/2}\right)_{0}n\kappa\right)_{1/2}$ state at
$n$, $\kappa$ and $Z$, which provide the minimum values of the
luminosity $L_{0}$.
% Results and discussion - IV.2
Table~\ref{tb:minimum_l0_res} corresponds to the case of resolved
opposite-parity states, whereas the case of unresolved states is
presented in Table~\ref{tb:minimum_l0}.
\begin{table}[h]
\begin{center}
\caption{
Cross section of the dielectronic recombination of a polarized electron
with He-like ion in the case of resolved levels ($R \geq 5$).
The electron energy is tuned in resonance with the
$\left(\left(1s2p_{1/2}\right)_{0}n\kappa\right)_{1/2}$ state.
Parameters $n$, $\kappa$ and $Z$ correspond to the
minimal value of luminosity $L_{0}$. $\Delta E =
E_{\left(\left(1s2p_{1/2}\right)_{0}n\kappa\right)_{1/2}} -
E_{\left(\left(1s2s\right)_{0}n\kappa\right)_{1/2}}$ is the energy
difference and $R$ is the coefficient indicating either the states
are resolved or not.
$\sigma_0$ is the cross section without the PNC effect and
$\sigma_{\text{PNC}}$ is the parity violating contribution.
$\Delta\sigma_{0}$ indicates the increase of the process cross
section related to the usage of Eq.~(\ref{eq:sigma_pnc}) instead
of Eq.~(\ref{eq:sigma_pnc_res}).
Notation: $y[x]$ represents
$y\times10^{x}$.
}
\label{tb:minimum_l0_res}
\end{center}

\begin{center}
\begin{tabular}{c|cccc|cccc}
\hline \hline
$Z$ & $n\kappa$ & $\Delta E$ (eV) & $R$ & $\varepsilon_{i}$(keV) & $L_{0}$(cm$^{-2}$s$^{-1}$) & $\sigma_{0}$(barn) & $\Delta \sigma_{0}$(\%) & $\sigma_{\text{PNC}}$(barn) \\
\hline
88 & $7s$ & 3.17(29) & 19.3 & 84.76 & 1.1[30] & 3.8[2] & 0.1\% & 1.2[-3]   \\
90 & $5s$ & 4.13(47) & 7.7 & 86.91 & 1.4[30] & 2.8[2] & 0.9\% & 9.1[-4]    \\
   & $6s$ & 2.51(47) & 7.9 & 88.36 & 5.3[29] & 2.7[2] & 0.9\% & 1.5[-3]    \\
   & $7s$ & 1.75(47) & 8.5 & 89.22 & 2.6[29] & 2.6[2] & 0.8\% & 2.0[-3]    \\
92 & $5s$ & 2.97(28) & 5.0 & 91.43 & 5.1[29] & 2.5[2] & 2.1\% & 1.4[-3]    \\
   & $7s$ & -1.60(28) & 7.2 & 93.86 & 1.5[29] & 2.2[2] & 1.1\% & -2.4[-3] \\
\hline \hline
\end{tabular}
\end{center}
\end{table}
\begin{table}[h]
\begin{center}
\caption{Cross section of the dielectronic recombination of a polarized
electron with He-like ion in the case of unresolved levels ($R < 5$).
The electron energy is tuned in resonance with the
$\left(\left(1s2p_{1/2}\right)_{0}n\kappa\right)_{1/2}$ state.
Parameters $n$, $\kappa$ and $Z$ correspond to the
minimal value of luminosity $L_{0}$.
$\Delta E = E_{\left(\left(1s2p_{1/2}\right)_{0}n\kappa\right)_{1/2}} -
E_{\left(\left(1s2s\right)_{0}n\kappa\right)_{1/2}}$ is the energy
difference and $R$ is the coefficient indicating either the states
are resolved or not.
$\sigma$ and $\overline{\sigma}$ are the cross sections
corresponding to the monoenergetic and non-monoenergetic energy
distribution of the incident electron beam, respectively.
$\sigma_0$ is the cross section without the PNC effect and
$\sigma_{\text{PNC}}$ is the parity violating contribution.
Notation: $y[x]$ represents $y\times10^{x}$.}
\label{tb:minimum_l0}
\end{center}
\begin{center}
\begin{tabular}{c|cccc|ccc|cc}
\hline \hline
$Z$ & $n\kappa$ & $\Delta E$(eV) & $R$ & $\varepsilon_{i}$(keV) & $L_{0}$(cm$^{-2}$s$^{-1}$) & $\sigma_{0}$(barn) & $\sigma_{\text{PNC}}$(barn) & $\overline{\sigma}_{0}$(barn eV) & $\overline{\sigma}_{\text{PNC}}$(barn eV)
\\ \hline
62 & $7s$        & -0.103(64)& 2.0 & 39.56 & 3.6[29] & 1.4[3] & -4.0[-3] & 4.8[2] &   5.5[-5] \\
88 & $7p_{1/2}$ & -2.46(29) & 4.4 & 84.76 & 1.7[30] & 2.8[1] & -2.6[-4] & 9.6[2] & -7.0[-6] \\
90 & $6p_{1/2}$ & -1.26(47) & 1.1 & 88.37 & 1.0[30] & 9.6[1] & -6.2[-4] & 1.6[3] & -2.6[-5] \\
92 & $6s$        & -1.07(28) & 3.0 & 92.96 & 7.3[28] & 2.5[2] & -3.8[-3] & 6.8[2] &   1.7[-5] \\
   & $6p_{1/2}$ & 2.38(27)   & 2.0 & 92.96 & 1.3[30] & 4.3[1] &   3.6[-4] & 1.6[3] & -1.8[-5] \\
   & $7p_{1/2}$ & 2.38(28)   & 3.2 & 93.86 & 8.1[29] & 2.8[1] &   3.8[-4] & 1.0[3] & -4.0[-6] \\
\hline \hline
\end{tabular}
\end{center}
\end{table}
%
%
%
% Results and discussion - IV.3
It is clearly seen from Table~\ref{tb:minimum_l0_res}, that the $R$
coefficient can be applied in order to distinguish cases of resolved and
unresolved states.
% Results and discussion - IV.4
Indeed, at the border value ($R = 5$), $\sigma_{0}$ increases only by
about 2\% for the calculations utilizing
Eq.~(\ref{eq:sigma_pnc}) instead of Eq.~(\ref{eq:sigma_pnc_res}).
% Results and discussion - IV.5
For other parameters $n$, $\kappa$ and $Z$ listed in
Table~\ref{tb:minimum_l0_res} the growth of the cross section amounts
to $1\%$ and less.
% ----------------------------------------------------------------------

%--------------------            5 paragraph       ------------------------
% Results and discussion - V.1
According to Tables~\ref{tb:minimum_l0_res}
and~\ref{tb:minimum_l0} the PNC effect seems to be most
promising for the dielectronic recombination of a polarized electron
with He-like uranium ($Z = 92$).
% Results and discussion - V.2
When the energy of the incident electron is tuned in the resonance
with the $\left(\left(1s 2p_{1/2}\right)_{0}6s\right)_{1/2}$
state, the ratio $\sigma_{\text{PNC}}/\sigma_{0}$ equals
$-1.5\times10^{-5}$.
% Results and discussion - V.3
After integration over $\varepsilon_{i}$ it turns into
$\overline{\sigma}_{\text{PNC}}/\overline{\sigma}_{0} = 2.5 \times 10^{-8}$.
% Results and discussion - V.4
Let us compare the obtained results with similar calculations
presented in Ref.~\cite{Gribakin_PRA72_032109:2005}.
% Results and discussion - V.5
In that work, the authors considered the process of the dielectronic
recombination into the $\left(2s\right)^{2}$ and $\left(2s
2p_{1/2}\right)_{0}$ states for $Z = 48$, where the enhancement of
the P-violating effect takes place due to the quasidegeneracy of
these levels.
% Results and discussion - V.6
The PNC asymmetry of the process considered in 
Ref.~\cite{Gribakin_PRA72_032109:2005} amounted to $5\times10^{-9}$,
while for the process considered in the present work it reaches
$1.5\times10^{-5}$.
% Results and discussion - V.7
The increase of the effect by more than three orders of magnitude is
caused by the fact that the admixing parameter $\xi$ for $Z = 48$, obtained
in Ref.~\cite{Gribakin_PRA72_032109:2005}, equals
$6.0\times10^{-9}$, whereas for Li-like uranium we get $\xi =
4.0\times10^{-6}$.
% ----------------------------------------------------------------------

\begin{figure}[h]
\caption{PNC cross sections of the dielectronic recombination into
the $\left(\left(1s 2p_{1/2}\right)_{0}7s\right)_{1/2}$ and
$\left(\left(1s 2s\right)_{0}7s\right)_{1/2}$ states of Li-like
samarium ($Z = 62$).
The difference $E_{i} - E_{\left(\left(1s
2p_{1/2}\right)_{0}7s\right)_{1/2}}$ determines uniquely the
energy of the incident electron.
The Solid line corresponds to $\sigma_{\text{PNC}}$, the dashed line is
the parity-violating contribution from the dielectronic recombination,
and the dotted line is the PNC contribution from the decay process
multiplied by a factor of 10.}
\includegraphics[trim=0 0 0 0, clip, width = 1.0\textwidth]
{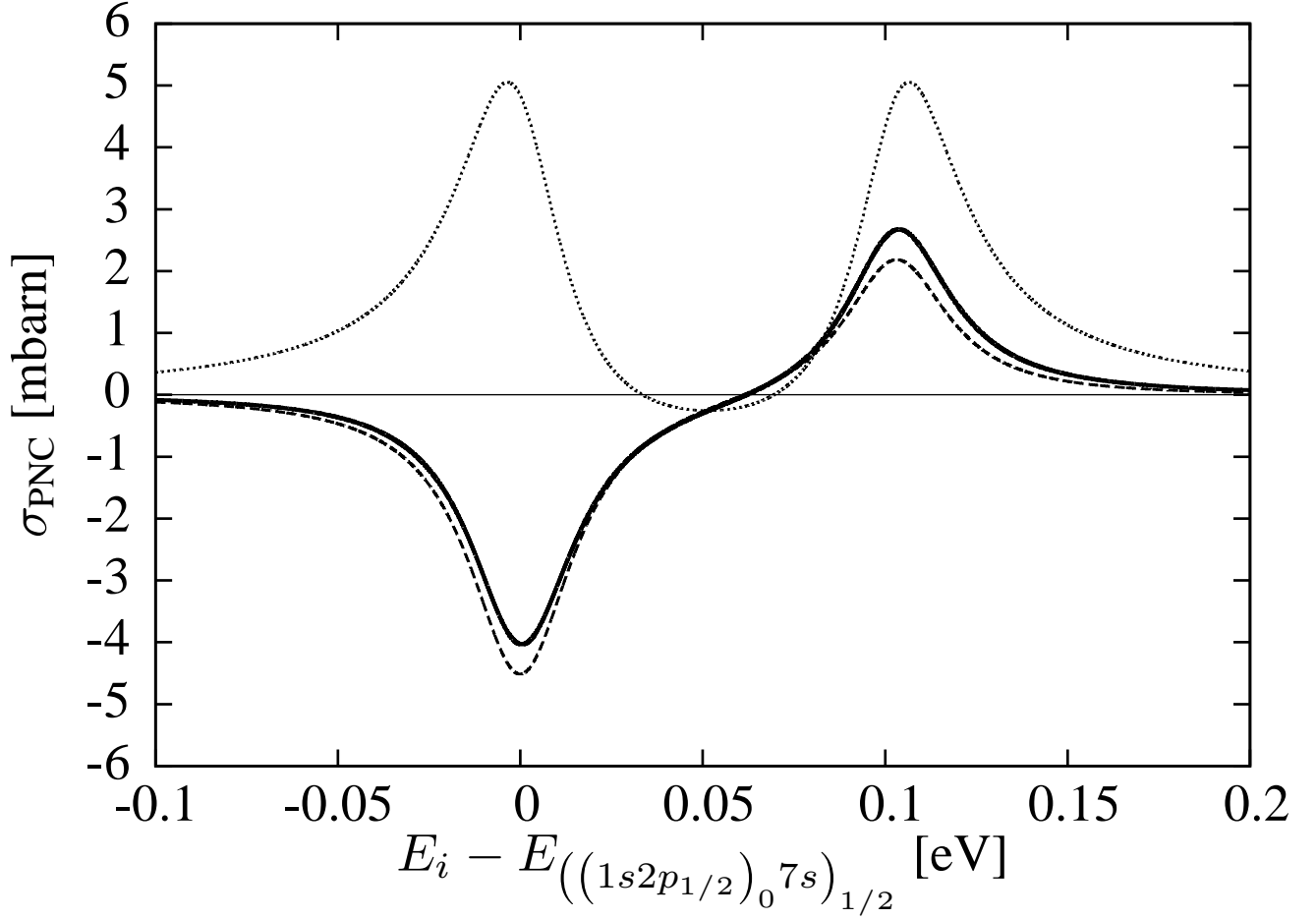} \label{ris:sigma_pnc}
\end{figure}

% --------------------             6 paragraph     ------------------------
% Results and discussion - VI.1
In Fig.~\ref{ris:sigma_pnc}, $\sigma_{\text{PNC}}$ is displayed as
a function of the energy of the incident electron in the case of
unresolved levels.
%
%
%
% Results and discussion - VI.2
As one can see from this figure, the PNC cross section is mainly
formed by the parity violation effect in the dielectronic
recombination process (third term in Eq.~(\ref{eq:sigma_pnc})).
% Results and discussion - VI.3
Nevertheless, the contribution from the subsequent radiative decay
(forth term in Eq.~(\ref{eq:sigma_pnc})) slightly enhances
$\sigma_{\text{PNC}}$ for the energy of the incident electron
tuned in resonance with $d_{2}$ state.
% Results and discussion - VI.4
Vice versa, for the energy tuned in resonance with the $d_{1}$
state a small decrease of the PNC contribution is observed.
% Results and discussion - VI.5
One can observe the energy of the incident electron at which
$\sigma_{\text{PNC}}$ turns to zero.
% Results and discussion - VI.6
It approximately corresponds to the energy just in between the
quasidegenerate $d_{1}$ and $d_{2}$ states.
% ----------------------------------------------------------------------

%-------------------            7 paragraph        ------------------------
% Results and discussion - VII.1
The experiment suggested in our paper involves a stored heavy ion
beam intersecting with a beam of polarized electrons in an
electron target or cooler.
% Results and discussion - VII.2
The polarized electrons can be produced with a semiconductor
photocathode with circularly polarized laser light
\cite{Pierce_APL26_670:1975}.
% Results and discussion - VII.3
They are electrostatically accelerated to the energies of tens of
keV that are required for the experiment.
% Results and discussion - VII.4
The cooler with a photocathode was constructed for instance for
the TSR storage ring at the MPI-K Heidelberg.
% Results and discussion - VII.5
Such coolers can in principle be made to produce polarized
electrons and they are now under consideration for the FAIR
facility and for the CRYRING at GSI / Darmstadt.
% ----------------------------------------------------------------------

%-------------------            8 paragraph        ------------------------
% Results and discussion - VIII.1
The high electron energy definition in the rest frame of the ion
is required to achieve the DR resonance.
% Results and discussion - VIII.2
The electron beam energy spread depends on the collision energy
$\varepsilon$ and the transverse $kT_{\perp}$ and the longitudinal
$kT_{||}$ temperatures of the electron beam: $\Delta \varepsilon=
\sqrt{\left(\ln 2 kT_{\perp}\right)^2+16\ln 2\varepsilon kT_{||}}$
\cite{Mueller_IJMS192_9:1999}.
% Results and discussion - VIII.3
The laser-produced beams of electrons are intrinsically cold and
can be further cooled using an adiabatic beam expansion technique.
% Results and discussion - VIII.4
Beams with a transverse temperature of 3.6~meV and a
longitudinal temperature of 38~$\mathrm{\mu eV}$ were produced in
electron cooler devices \cite{Lestinsky_AJ698_648:2009}.
% Results and discussion - VIII.5
Accordingly, the energy spread of a few eV at 90 keV can
be experimentally achieved at present.
% Results and discussion - VIII.6
This means that the DR resonance structure will be integrated out.
% and the PNC effect of the order of $10^{-8}$ can be observed.
% Results and discussion - VIII.7
To take full advantage of the enhancement of PNC in
dielectronic recombination the electron energy spread must be made
smaller than 0.1 eV at 90 keV, which is nowadays not possible.
% Results and discussion - VIII.8
Therefore further developments will be required to produce the
electron beam that is cold enough.
% Results and discussion - VIII.9
In addition, the ion beam momentum spread should be reduced below
$10^{-6}$.
% Results and discussion - VIII.10
This, however, was demonstrated at the storage ring ESR albeit
with a significant reduction of the beam intensity
\cite{Steck_PRL77_3803:1996}.

%-------------------            9 paragraph        ------------------------
% Results and discussion - IX.1
%Besides all of the above the considered process is masked
%by the direct radiative recombination cross section.
% Results and discussion - IX.2
%Nevertheless, we hope that the presented calculations will help to
%find more feasible scenarios of P-violating experiments in the
%case of heavy highly-charged ions.
% ----------------------------------------------------------------------

% ==========­=====        END RESULT AND DISCUSSION      ==========­========

% ==========­=====               CONCLUSION                ==========­========
\section{CONCLUSION}
%-------------------             I paragraph              ------------------
% Conclusion - I.1
In the present work we have considered the PNC effect on the cross
section of the dielectronic recombination into the $\left( \left(
1s 2p_{1/2} \right)_{0} n\kappa \right)_{1/2}$ and $\left( \left(
1s 2s \right)_{0} n\kappa \right)_{1/2}$ states of heavy Li-like
ions.
% Conclusion - I.2
The calculations have been performed for the parameters $n$, $\kappa$ and $Z$ which provide the enhancement of the P-violation effect due to quasidegeneracy of the corresponding levels.
% Conclusion - I.3
It has been found that at energies of incident electron tuned in resonance with the
$\left(\left(1s 2p_{1/2}\right)_{0}n\kappa\right)_{1/2}$ state the PNC effect becomes most pronounced.
% Conclusion - I.4
The estimation of the PNC asymmetry for the most promising case of
$\left(\left(1s 2p_{1/2}\right)_{0} 6s\right)_{1/2}$ and
$\left(\left(1s 2s\right)_{0} 6s\right)_{1/2}$ states for $Z = 92$ has given
$-1.5 \times 10^{-5}$, which is by several orders of magnitude bigger than
the result obtained for a similar process in Ref.~\cite{Gribakin_PRA72_032109:2005}.
%Non-monoenergetic electron beam has been also studied.
%In this case the integration over the energy of the incident electron
%has been performed analytically.
% ----------------------------------------------------------------------
% ==========­==========­==       END CONCLUSION            ==========­========

% ==========­==========­==       ACKNOWLEDGEMENTS         ==========­========
\section{ACKNOWLEDGEMENTS}
% ----------------------         I paragraph             ------------------
% Acknowledgements - I.1
Fruitful discussions with M.~G.~Kozlov are gratefully acknowledged.
% Acknowledgements - I.2
The work reported in this paper was supported by RFBR (Grants No. 12-02-31133 and 13-02-00630), by DFG, by GSI, by DAAD, and by the Ministry of Education and Science of Russian Federation (Program ''Scientific and scientific-pedagogical personnel of innovative Russia'', Grant No. 8420).
% Acknowledgements - I.3
The work of VAZ was supported by the German-Russian Interdisciplinary Science Center (G-RISC) and by the FAIR Russia Research Center.
% Acknowledgements - I.4
AVM acknowledges financial support by the ''Dynasty'' foundation.
% ----------------------------------------------------------------------
% ==========­==========­=        END ACKNOWLEDGEMENTS        ==========­=====

%%%%%%%%%%­%%%%%%%%%%­%%%%%%%%%%­%%%%%%%%%%­%%%%%%%%%%­%%%%%%%%%%­%%%%%%%%%%­%%%%%%
%\pagebreak

%
%\pagebreak

%%%%%%%%%%­%%%%%%%%%%­%%%%%%%%%%­%%%%%%%%%%­%%%%%%%%%%­%%%%%%%%%%­%%%%%%%%%%­%%%%%%%

\begin{thebibliography}{99}
%%%%%%%%%%­%%%%%%%%%%­%%%%%%%%%%­%%%%%%%%%%­%%%%%%%%%%­%%%%%%%%%%­%%%%%%%%%%­%%%%%%
% 1
\bibitem{Khriplovich}
I.~B. ~Khriplovich,
\textit{Parity Nonconservation in Atomic Phenomena} (Gordon and Breach, London, 1991).
% 2
\bibitem{Khriplovich_PST112_52:2004}
I.~B.~Khriplovich,
Phys. Scr. \textbf{T112}, 52 (2004).
% 3
\bibitem{Ginges_PR397_63:2004}
J.~S.~M.~Ginges and   V.~V.~Flambaum,
Phys. Rep. {\bf 397}, 63 (2004).
% 4
\bibitem{Wood_S275_1759:1997}
C.~S.~Wood, S.~C.~Bennett, D.~Cho, B.~P.~Masterson, J.~L.~Roberts, C.~E.~Tanner, and C.~E.~Wieman,
Science \textbf{275}, 1759 (1997).
% 5
\bibitem{Bennett_PRL82_2484:1999}
S.~C.~Bennett and C.~E.~Wieman,
Phys. Rev. Lett. \textbf{82}, 2484 (1999);
Phys. Rev. Lett. \textbf{83}, 889 (1999).
% 6
\bibitem{Shabaev_PRL94_213002:2005}
V.~M.~Shabaev, K.~Pachucki, I.~I.~Tupitsyn, and V.~A.~Yerokhin,
Phys. Rev. Lett. \textbf{94}, 213002 (2005);
V.~M.~Shabaev, I.~I.~Tupitsyn, K.~Pachucki, G.~Plunien, and V.~A.~Yerokhin,
Phys. Rev. A \textbf{72}, 062105 (2005).
% 7
\bibitem{Porsev_PRL102_181601:2009}
S.~G.~Porsev, K.~Beloy, and A.~Derevianko,
Phys. Rev. Lett. \textbf{102}, 181601 (2009).
% 8
\bibitem{Dzuba_PRL109_203003:2012}
V.~A.~Dzuba, J.~C.~Berengut, V.~V.~Flambaum, and B.~Roberts,
Phys. Rev. Lett. {\bf 109}, 203003 (2012).
% 9
\bibitem{Gorshkov_JL19_394:1974}
V.~G.~Gorshkov and L.~N.~Labzowsky,
Zh.~Eksp.~Teor.~Fiz.~Pis'ma,
\textbf{19}, 768 (1974)
[JETP Lett., \textbf{19}, 394 (1974)];
Zh.~Eksp.~Teor.~Fiz., \textbf{69}, 1141 (1975)
[Sov. Phys. JETP,
\textbf{42}, 581 (1975)].
% 10
\bibitem{Schafer_PRA40_7362:1989}
A.~Sch\"afer, G.~Soff, P.~Indelicato, B.~M\"uller, and W.~Greiner,
Phys. Rev. A {\bf 40}, 7362 (1989).
% 11
\bibitem{Pindzola_PRA47_4856:1993}
M.~S.~Pindzola, Phys. Rev. A \textbf{47}, 4856 (1993).
% 12
\bibitem{dun96}
R.W. Dunford, Phys. Rev. A {\bf 54}, 3820 (1996).
\bibitem{Labzowsky_PRA63_054105:2001}
L.~N.~Labzowsky, A.~V.~Nefiodov, G.~Plunien, G.~Soff, R.~Marrus, and D.~Liesen,
Phys. Rev. A {\bf 63}, 054105 (2001).
% 13
\bibitem{Gribakin_PRA72_032109:2005}
G.~F.~Gribakin, F.~J.~Currel, M.~G.~Kozlov, and A.~I.~Mikhailov,
Phys. Rev. A \textbf{72}, 032109 (2005); Phys. Rev. A \textbf{80},
049901(E) (2009); arXiv:physics/0504129 (2005).
% 14
\bibitem{Maiorova_JPB42_205002:2009}
A.~V.~Maiorova, O.~I.~Pavlova, V.~M.~Shabaev, C.~Kozhuharov, G.~Plunien, and T.~St\"ohlker,
J. Phys. B: At. Mol. Opt. Phys. {\bf 42}, 205002 (2009).
% 15
\bibitem{Shabaev_PRA81_052102:2010}
V.~M.~Shabaev, A.~V.~Volotka, C.~Kozhuharov, G.~Plunien, and T.~St\"ohlker,
Phys. Rev. A {\bf 81}, 052102 (2010).
% 16
\bibitem{Ferro_PRA81_062503:2010}
F.~Ferro, A.~Artemyev, T.~St\"ohlker, and A.~Surzhykov, Phys. Rev.
A {\bf 81}, 062503 (2010).
% 17
\bibitem{Ferro_PRA83_052518:2011}
F.~Ferro, A.~Surzhykov, and T.~St\"ohlker, Phys. Rev. A {\bf 83},
052518 (2011).
% 18
\bibitem{Maiorova_JPB44_225003:2011}
A.~V.~Maiorova, V.~M.~Shabaev, A.~V.~Volotka, V.~A.~Zaytsev,
G.~Plunien, and T.~St\"ohlker, J. Phys. B: At. Mol. Opt. Phys.
{\bf 44}, 225003 (2011).
% 19
\bibitem{Gunst_PRA87_032714:2013}
J.~Gunst, A.~Surzhykov, A.~Artemyev, S.~Fritzsche, S.~Tashenov, A.~Maiorova, V.~M.~Shabaev, and T.~St\"ohlker,
Phys. Rev. A {\bf 87}, 032714 (2013).
% 20
\bibitem{Shabaev_PRA50_4521:1994}
V.~M.~Shabaev,
Phys. Rev. A {\bf 50}, 4521 (1994).
% 21
\bibitem{Shabaev_PR356_119:2002}
V.~M.~Shabaev,
Phys. Rep. \textbf{356}, 119 (2002).
% 22
\bibitem{Rose}
M.~E.~Rose,
\textit{Elementary Theory of Angular Momentum} (Wiley, New York,   1957).
% 23
\bibitem{Eichler_PR439_1:2007}
J.~Eichler and T.~St\"ohlker,
Phys. Rep. {\bf 439}, 1 (2007).
% 24
%\bibitem{Salvat_CPC90_151:1995}
%F.~Salvat, J.~M.~Fernandez-Varea, and Jr.~W.~Williamson,
%Comput. Phys. Commun. \textbf{90}, 151 (1995).
% 24
\bibitem{Zaytsev_PS}
V.~A.~Zaytsev, A.~V.~Maiorova, V.~M.~Shabaev, A.~V.~Volotka, and G.~Plunien,
Phys. Scr. {\bf T156}, 014028 (2013).
% 25
\bibitem{Artemyev_PRA71_062104:2005}
A.~N.~Artemyev, V.~M.~Shabaev, V.~A.~Yerokhin, G.~Plunien, and G.~Soff,
Phys. Rev. A \textbf{71}, 062104 (2005).
% 26
\bibitem{Pierce_APL26_670:1975}
D.~T.~Pierce, F.~Meier, and P.~Z\"urcher,   App. Phys. Lett.
\textbf{26}, 670-672 (1975).
% 27
\bibitem{Mueller_IJMS192_9:1999}
A.~M\"uller, Int. J. Mass Spectrom. \textbf{192}, 9 (1999).
% 28
\bibitem{Lestinsky_AJ698_648:2009}
M.~Lestinsky, N.~R.~Badnell, D.~Bernhardt, M.~Grieser,
J.~Hoffmann, D.~Luki\'c, A.~M\"uller, D.~A.~Orlov, R.~Repnow,
D.~W.~Savin, E.~W.~Schmidt, M.~Schnell, S.~Schippers, A.~Wolf, and
D.~Yu, Astr. J. \textbf{698}, 648 (2009).
% 29
\bibitem{Steck_PRL77_3803:1996}
M.~Steck, K.~Beckert, H.~Eickhoff, B.~Franzke, F.~Nolden,
H.~Reich, B.~Schlitt, and T.~Winkler, Phys. Rev. Lett.
\textbf{77}, 3803 (1996).

\end{thebibliography}
\end {document}